\documentclass[preprint,showpacs,preprintnumbers,amsmath,amssymb]{revtex4}
\usepackage{amsmath}

\usepackage{graphicx}
\usepackage{dcolumn}
\usepackage{bm}
\usepackage{amsmath} 

\newcommand{\qed}{\nobreak \ifvmode \relax \else
      \ifdim\lastskip<1.5em \hskip-\lastskip
      \hskip1.5em plus0em minus0.5em \fi \nobreak
      \vrule height0.75em width0.5em depth0.25em\fi}

\begin{document}

\preprint{}

\title{Hypergeometric/Difference-Equation-Based Separability Probability Formulas and Their Asymptotics for Generalized Two-Qubit States Endowed with Random Induced Measure}
\author{Paul B. Slater}
 \email{slater@kitp.ucsb.edu}
\affiliation{%
University of California, Santa Barbara, CA 93106-4030\\
}
\date{\today}
            
\begin{abstract}
We find equivalent hypergeometric- and difference-equation-based formulas, $Q(k,\alpha)= G_1^k(\alpha) G_2^k(\alpha)$, for $k = -1, 0, 1,\ldots,9$, for that (rational-valued) portion of the total separability probability 
for generalized two-qubit states endowed with random induced measure, for which   the determinantal inequality $|\rho^{PT}| >|\rho|$ holds. Here $\rho$ denotes a $4 \times 4$ density matrix and $\rho^{PT}$, its partial transpose, while $\alpha$ is a Dyson-index-like parameter with $\alpha = 1$ for the standard (15-dimensional) convex set of two-qubit states. The dimension of the space in which these density matrices is embedded is $4 \times (4 +k)$. For the symmetric case of $k=0$, we obtain the previously reported Hilbert-Schmidt formulas, with
(the two-re[al]bit case) $Q(0,\frac{1}{2}) = \frac{29}{128}$, (the standard two-qubit case) $Q(0,1)=\frac{4}{33}$, and 
(the two-quater[nionic]bit case) $Q(0,2)= \frac{13}{323}$. The factors $G_2^k(\alpha)$ can be written as the sum of weighted hypergeometric functions $_{p}F_{p-1}$, $p \geq 7$, 
all with argument $\frac{27}{64} =(\frac{3}{4})^3$. 
We find formulas for the upper and lower parameter sets of these functions and, then, equivalently express $G_2^k(\alpha)$ in terms of first-order difference equations. The factors $G_1^k(\alpha)$ are equal to $(\frac{27}{64})^{\alpha-1}$ times ratios of products of six Pochhammer symbols involving the indicated parameters. Some remarkable $\alpha-$ and $k$-specific invariant asymptotic
properties (again, involving $\frac{27}{64}$ and related quantities) of  separability probability formulas emerge.
\end{abstract}


\maketitle
\section{Introduction}
In a previous paper \cite{LatestCollaboration}, a family of ($\alpha$-specific) formulas was obtained for the (total) separability probabilities of generalized two-qubit states. Here, we examine a related quantity informing us of that portion of the separability probabilities associated with the determinantal inequality $|\rho^{PT}| > |\rho|$. Here, $\rho$ denotes a $4 \times 4$ density matrix and $\rho^{PT}$, its partial transpose, with $\alpha$ serving as a Dyson-index-like parameter. Of course, by the Peres-Horodecki conditions \cite{asher,michal}, a necessary and sufficient condition for separability in this $4 \times 4$ case is that 
$|\rho^{PT}|>0$, while $|\rho| \geq 0$ itself certainly holds, independently of any separability considerations. So, the total separability probability can clearly be expressed as the sum of that part for which $|\rho^{PT}| > |\rho|$ and that for 
which $|\rho| > |\rho^{PT}| \geq  0$. The former part will be the one of immediate concern here.

To obtain the new formulas to be reported, we employ the Legendre-polynomial-based density approximation (Mathematica-implemented) algorithm of Provost 
\cite{Provost}, utilizing the previously-obtained moment 
formula \cite[sec. II]{WholeHalf} (cf. \cite{Beran})
\begin{align*}
\left\langle \left\vert \rho\right\vert
^{k}\left(  \left\vert \rho^{PT}\right\vert -\left\vert \rho\right\vert
\right)  ^{n}\right\rangle /\left\langle \left\vert \rho\right\vert
^{k}\right\rangle   &  =
\left(  -1\right)  ^{n}\frac{\left(  \alpha\right)  _{n}\left(  \alpha
+\frac{1}{2}\right)  _{n}\left(  n+2k+2+5\alpha\right)  _{n}}{2^{4n}\left(
k+3\alpha+\frac{3}{2}\right)  _{n}\left(  2k+6\alpha+\frac{5}{2}\right)
_{2n}}\\
& \times~_{4}F_{3}\left(
\genfrac{}{}{0pt}{}{-\frac{n}{2},\frac{1-n}{2},k+1+\alpha,k+1+2\alpha
}{1-n-\alpha,\frac{1}{2}-n-\alpha,n+2k+2+5\alpha}%
;1\right)  ,
\end{align*}
while in \cite{LatestCollaboration}, the moment formula \cite[sec. X.D.6]{MomentBased}
\begin{gather*} \label{nequalzero}
\left\langle \left\vert \rho^{PT}\right\vert ^{n}\right\rangle =\frac
{n!\left(  \alpha+1\right)  _{n}\left(  2\alpha+1\right)  _{n}}{2^{6n}\left(
3\alpha+\frac{3}{2}\right)  _{n}\left(  6\alpha+\frac{5}{2}\right)  _{2n}}\\
+\frac{\left(  -2n-1-5\alpha\right)  _{n}\left(  \alpha\right)  _{n}\left(
\alpha+\frac{1}{2}\right)  _{n}}{2^{4n}\left(  3\alpha+\frac{3}{2}\right)
_{n}\left(  6\alpha+\frac{5}{2}\right)  _{2n}}~_{5}F_{4}\left(
\genfrac{}{}{0pt}{}{-\frac{n-2}{2},-\frac{n-1}{2},-n,\alpha+1,2\alpha
+1}{1-n,n+2+5\alpha,1-n-\alpha,\frac{1}{2}-n-\alpha}%
;1\right) 
\end{gather*}
had been utilized for the density-approximation purposes there. (These [random-induced measure \cite{aubrun2}] moment formulas had been developed based on calculations solely for the two-rebit [$\alpha=\frac{1}{2}$] and two-qubit [$\alpha=1$] cases. However, they do appear, as well, remarkably, to apply to the two-quater[nionic]bit [$\alpha =2$] case  \cite{FeiJoynt}. No explicit formal extension of the Peres-Horodecki partial-transposition conditions \cite{asher,michal} to two-quaterbit systems seems to have been developed, however [cf. \cite{carl,aslaksen,asher2}].) 

In \cite{LatestCollaboration}, $\alpha$-specific formulas ($\alpha = 1,2,\ldots,13$ and $\frac{1}{2}, \frac{3}{2},\frac{5}{2},\frac{7}{2}$) as a function of $k$ for the total 
($|\rho^{PT}| >0$) separability probabilities had been derived. Here,  contrastingly, we will find $k$-specific formulas 
($k=-1,0,1,\ldots,9$) as a function of $\alpha$ for the indicated one ($|\rho^{PT}| > |\rho|$) of their two component parts.
We utilize an exceptionally large number (15,801) number of moments in the routine of Provost \cite{Provost}, helping to reveal--to extraordinarily
high accuracy--the rational values that the corresponding separability  probabilities strongly appear to assume.
Sequences ($\alpha =1, 2,\ldots,30,\ldots$) of such rational values, then, serve as input to the FindSequenceFunction command of Mathematica to obtain the initial set of $k$-specific (hypergeometric-based) formulas for $Q(k,\alpha)$, which we, then, further manipulate.
\section{Common features of the $k$-specific formulas}
For each $k = -1, 0, 1,\ldots,9$, the FindSequenceFunction command yields what we can
consider as a large, rather cumbersome (several-page) formula, which we denote by $Q(k,\alpha)$. It, in fact, faithfully reproduces the inputted rational-valued (separability probability) sequences. This fidelity is indicated by numerical calculations to apparently arbitrarily high accuracy (hundreds of digits). (The difference equation results below [sec.~\ref{Equivalent}] will provide a basis for our observation as to the rational-valuedness of the separability probabilities.)

\begin{figure}
\includegraphics{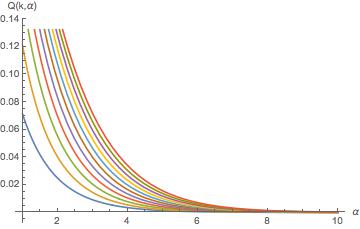}
\caption{\label{fig:Raw}Plots of $Q(k,\alpha)$ over the range
 $\alpha \in [1,10]$, for $k=-1,\ldots,9$. For fixed $\alpha$, we have $Q(k_1,\alpha) > Q(k_2,\alpha)$, if $k_1>k_2$.}
\end{figure}

\begin{figure}
\includegraphics{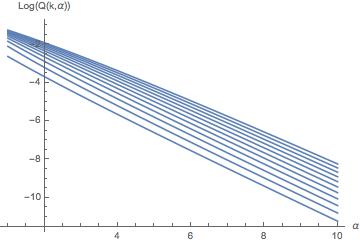}
\caption{\label{fig:Log}Plots of $\log{Q(k,\alpha)}$ over the range
 $\alpha \in [1,10]$, for $k=-1,\ldots,9$. For fixed $\alpha$, $\log{Q(k_1,\alpha)} > \log{Q(k_2,\alpha)}$, if $k_1>k_2$.}
\end{figure}

In Fig.~\ref{fig:Raw}, we show plots of $Q(k,\alpha)$ over the range
 $\alpha \in [1,10]$, for $k=-1,\ldots,9$. For fixed $\alpha$, we have $Q(k_1,\alpha) > Q(k_2,\alpha)$, if $k_1>k_2$. In Fig.~\ref{fig:Log}, we show a parallel plot, exhibiting linear-like behavior, for $\log{Q(k,\alpha)}$.
\subsection{Distinguished $_{7}F_{6}$ function with 2 as an upper parameter} \label{First7F6}
In each of the eleven $k$-specific formulas obtained, there is a distinguished $_{7}F_{6}$ function, with
the (omnipresent) argument of  $\frac{27}{64} =(\frac{3}{4})^3$ (cf. \cite{guillera} \cite[Ex. 8.6, p. 159]{koepf}), 
with 2 as one of the seven upper parameters (cf. \cite{slaterJModPhys}). The lower (bottom) six parameters
conform to the simple linear rule
\begin{equation} \label{bottom}
\{b_1,b_2,b_3,b_4,b_5,b_6\}=
\end{equation}
\begin{displaymath}
\left\{\alpha +\frac{2 k}{5}+\frac{23}{10},\alpha +\frac{2 k}{5}+\frac{5}{2},\alpha
   +\frac{2 k}{5}+\frac{27}{10},\alpha +\frac{2 k}{5}+\frac{29}{10},\alpha +\frac{2
   k}{5}+\frac{31}{10},\alpha +k+3\right\}.
\end{displaymath}
The six  upper parameters (aside from the 2 mentioned) can be broken into one set of two (summing to an integer), incorporating consecutive fractions having 6's in their denominators, 
and one set of four (also summing to an integer), incorporating consecutive fractions having 5's in their denominators. 

For the set of two, the smaller of the two entries abides by the rule
\begin{equation} \label{upper1}
u_1 = \frac{1}{6} \left(6 \alpha +4 \left\lfloor \frac{k}{3}\right\rfloor +2 \left\lfloor
   \frac{k+1}{3}\right\rfloor +11\right),
\end{equation}
where the (integer-valued) floor function is employed,
and the larger entry by
\begin{equation} \label{upper2}
u_2= \frac{1}{6} \left(6 \alpha +2 \left\lfloor \frac{k}{3}\right\rfloor +4 \left\lfloor
   \frac{k+1}{3}\right\rfloor +13\right).
\end{equation}
For $k = 1$, for illustrative purposes, application of these two rules yields $\left\{\alpha +\frac{11}{6},\alpha +\frac{13}{6}\right\}$, and 
for $k=5$, we have $\left\{\alpha +\frac{19}{6},\alpha +\frac{23}{6}\right\}$. (We have that $u_1+u_2$ is an integer. The sequence of those integers is reproduced in A004523 
[``Two even followed by one odd"] and 
A232007 [``Maximal number of moves needed to reach every square by a knight from a fixed position on an n X n chessboard, or -1 if it is not possible to reach every square"] in the On-Line Encyclopedia of Integer Sequences [https://oeis.org/ol.html].)

For the complementary set of four upper parameters, 
the entries in order of increasing magnitude are expressible as 
\begin{equation} \label{upper3}
u_3=\alpha+\frac{1}{5} \left(3 \left\lfloor \frac{k-4}{5}\right\rfloor +2 \left\lfloor
   \frac{k-3}{5}\right\rfloor +2 \left\lfloor \frac{k-2}{5}\right\rfloor +3 \left\lfloor
   \frac{k-1}{5}\right\rfloor +16\right),
\end{equation}
\begin{displaymath}
u_4=\alpha +\frac{1}{5} \left(3 \left\lfloor \frac{k-4}{5}\right\rfloor +2 \left\lfloor
   \frac{k-3}{5}\right\rfloor +\left\lfloor \frac{k-2}{5}\right\rfloor +4 \left\lfloor
   \frac{k-1}{5}\right\rfloor +17\right),
\end{displaymath}
\begin{displaymath}
u_5=\alpha +\frac{1}{5} \left(2 \left\lfloor \frac{k-4}{5}\right\rfloor +3 \left\lfloor
   \frac{k-3}{5}\right\rfloor +\left\lfloor \frac{k-2}{5}\right\rfloor +4 \left\lfloor
   \frac{k-1}{5}\right\rfloor +18\right),
\end{displaymath}
and
\begin{displaymath}
u_6=\alpha +\frac{1}{5} \left(2 \left\lfloor \frac{k-4}{5}\right\rfloor +3 \left\lfloor
   \frac{k-3}{5}\right\rfloor +\left\lfloor \frac{k-2}{5}\right\rfloor +4 \left\lfloor
   \frac{k-1}{5}\right\rfloor +19\right).
\end{displaymath}
For $k = 1$, for illustrative purposes, application of these four rules yields $\left\{\alpha +\frac{9}{5},\alpha +\frac{11}{5},\alpha +\frac{12}{5},\alpha
   +\frac{13}{5}\right\}$, and 
for $k=5$, we have $\left\{\alpha +\frac{16}{5},\alpha +\frac{17}{5},\alpha +\frac{18}{5},\alpha
   +\frac{19}{5}\right\}$.
\subsection{Distinguished $_{7}F_{6}$ function with 1 as an upper parameter} \label{Isolated}
Each $k$-specific formula $Q(k,\alpha)$ we have found 
also incorporates a second 
$_{7}F_{6}$ function (again with argument $\frac{27}{64}$, which is, to repeat, invariably the case throughout this paper), having all its thirteen parameters equalling 1 less those in the function just described. 
(A basic transformation exists 
[consulting the HYP manual of C. Krattenthaler, available at 
\newline www.mat.univie.ac.at, allowing one to convert the thirteen [twelve  $\alpha$-dependent parameters, plus 1]  of this $_7F_6$ function [that is, add 1 to each of them] to those of the other 
$_7F_6$ one first described.)

\subsection{The remaining $_{p}F_{p-1}$ functions, all with $p>7$.}
Now, all the remaining $m$ hypergeometric functions yielded by
the FindSequenceFunction command for each of the $k$-specific cases possess, to begin with, 
the same seven upper parameter (2 plus those indicated in 
(\ref{upper1}), (\ref{upper2}) and (\ref{upper3})) and the same six lower parameters (\ref{bottom}), as in the first $_{7}F_{6}$ function detailed above 
(sec.~\ref{First7F6}). Then, the seven upper parameters are supplemented by from one to $m$ 2's, and the six lower parameters supplemented by  from 1 to $m$ 1's.

From $k = -1$ to $k=9$, the eleven observed values of $m$ are
\begin{equation} \label{numbers}
\{m_{-1},m_0,m_1,m_2,m_3,m_4,m_5,m_6,m_7,m_8,m_9\}=\{3, 5, 5, 6, 6, 7, 9, 8, 10, 10, 10\}.
\end{equation}
\subsection{Large $\alpha$-free terms collapsing to 0}
We now point out a rather remarkable property of the formulas yielded by the \newline FindSequenceFunction command. If we isolate those (often quite bulky) terms that do not involve any of the hypergeometric functions described above, we find (to hundreds of digits of accuracy) that they collapse to zero. These terms, typically, do contain hypergeometric functions similar in nature to those described above, but with the crucial difference that the Dyson-index-like parameter $\alpha$ does {\it not}
occur among their upper and lower parameters. Thus, we are left with formulas $Q(k,\alpha)$ that are simply sums of  $m_k+2$ weighted $_pF_{p-1}$ functions (of $\alpha$), $p=7,\ldots,7+m_k$.
\section{Decomposition of $Q(k,\alpha)$  into the product $G_1^k(\alpha) G_2^k(\alpha)$}
The formulas we have obtained $Q(k,\alpha)$ can all be written--we have found--in the product form $G_1^k(\alpha) G_2^k(\alpha)$. The  $G_2^k(\alpha)$ factor involves the summation of the hypergeometric functions $_{p}F_{p-1}$ indicated above, each such function weighted by a polynomial in $\alpha$, the degrees of the polynomials diminishing as $p$ increases. Let us first analyze the other (hypergeometric-free) factor $G_1^k(\alpha)$, primarily
involving ratios of products of gamma functions.
\subsection{Hypergeometric-function-independent factor $G_1^k(\alpha)$}
Some supplementary computations (involving an independent use of the FindSequenceFunction command) indicated that this 
(hypergeometric-free) factor might be written quite concisely as
\begin{equation}
G_1^k(\alpha)= (\frac{27}{64})^{\alpha-1} \frac{\left(u_1\right)_{\alpha -1} \left(u_2\right)_{\alpha -1} \left(u_3\right)_{\alpha
   -1} \left(u_4\right)_{\alpha -1} \left(u_5\right)_{\alpha -1} \left(u_6\right)_{\alpha
   -1}}{\left(b_1\right)_{\alpha -1} \left(b_2\right)_{\alpha -1} \left(b_3\right)_{\alpha
   -1} \left(b_4\right)_{\alpha -1} \left(b_5\right)_{\alpha -1} \left(b_6\right)_{\alpha
   -1}},
\end{equation}
where the Pochhammer symbol (rising factorial) is employed. Note that
$G_1^k(1)=1$.
\subsection{Hypergeometric-function-dependent factor $G_2^k(\alpha)$}
\subsubsection{Canonical form}
\begin{figure}
\includegraphics[scale=0.95,trim=2cm 8cm 4cm 2cm,clip]{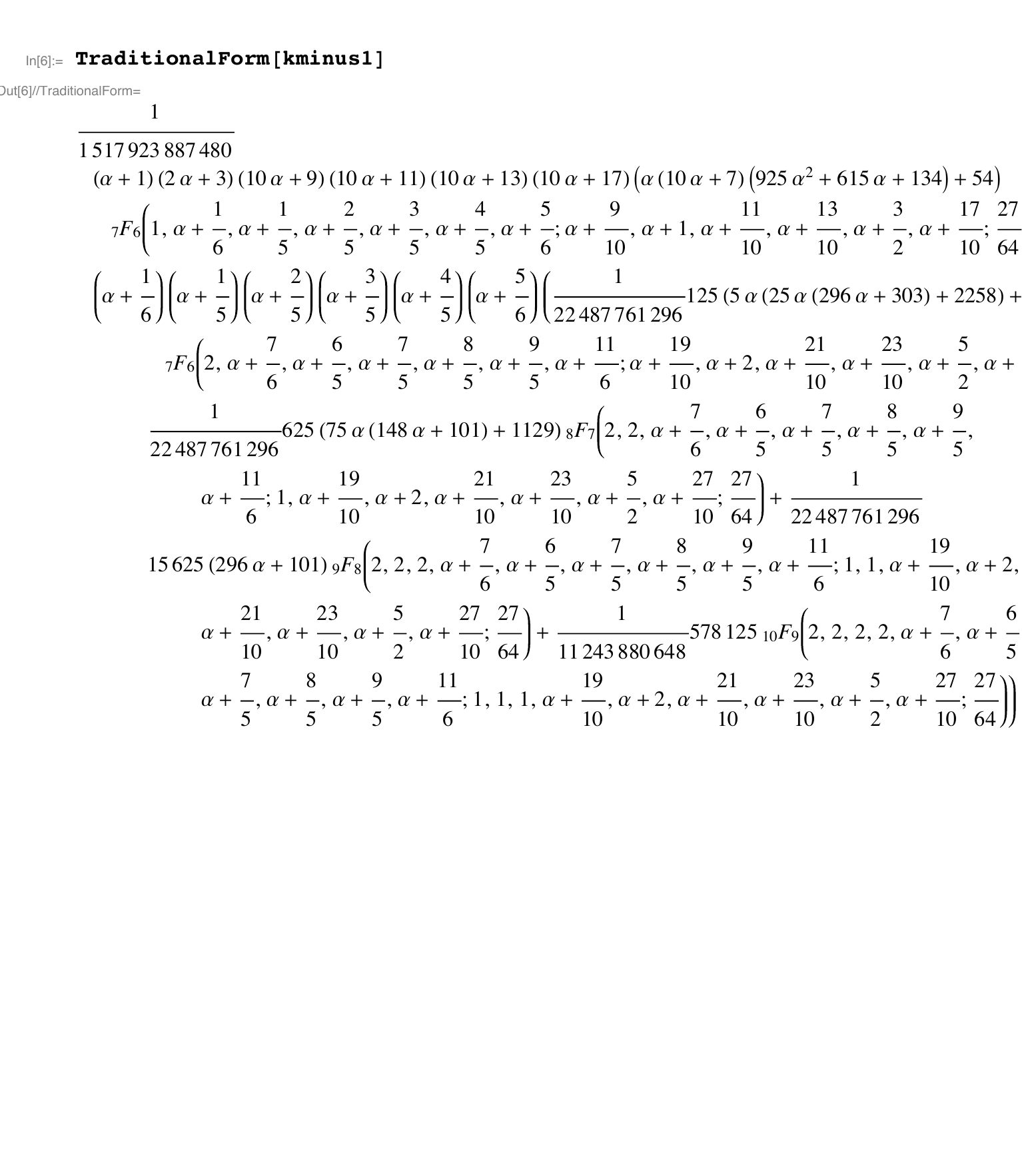}
\caption{\label{fig:kminus1}$G_2^{-1}(\alpha)$}
\end{figure}

\begin{figure}
\includegraphics[scale=0.95]{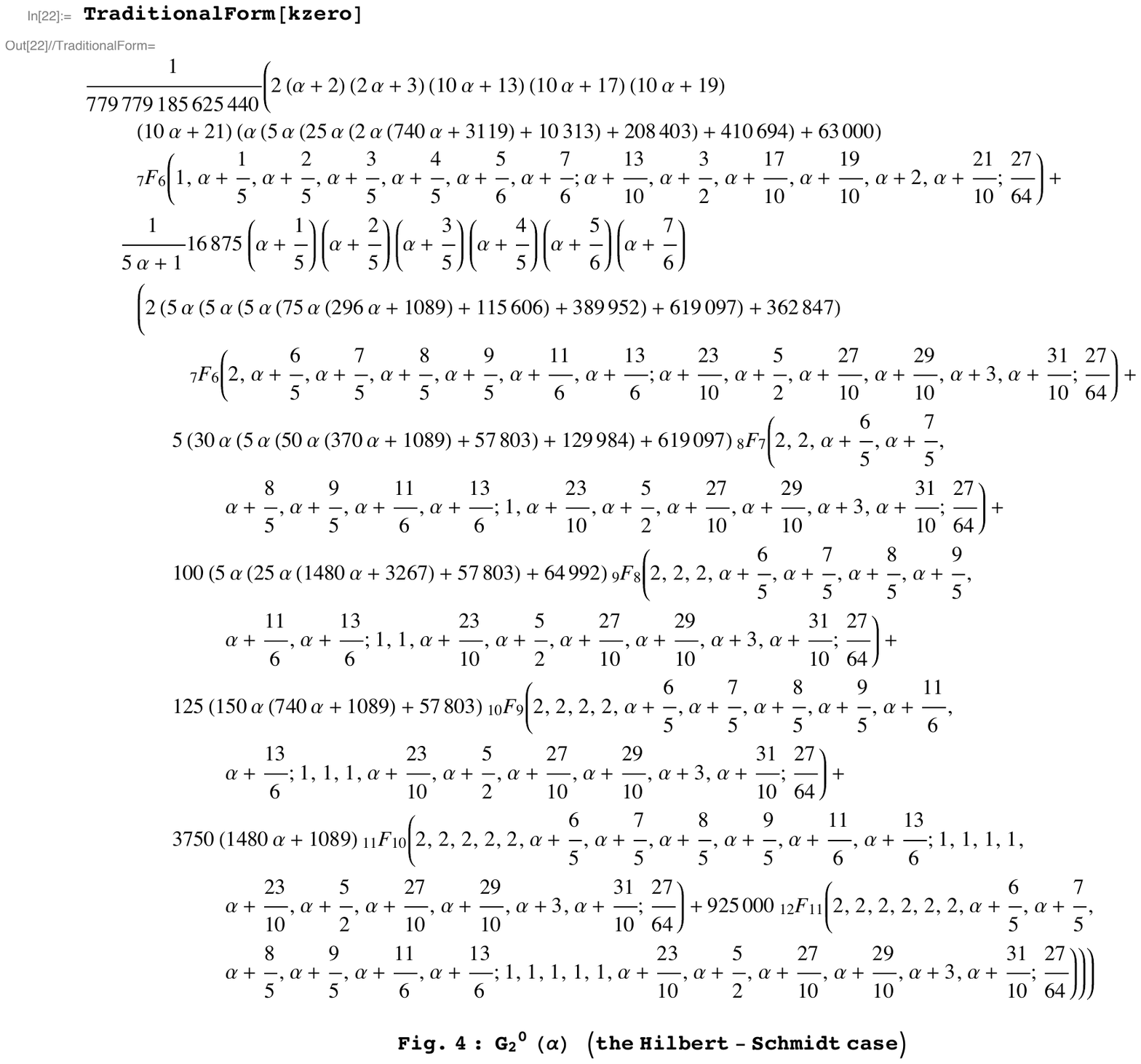}
\caption{\label{fig:zero}$G_2^{0}(\alpha)$, the Hilbert-Schmidt case}
\end{figure}

\begin{figure}
\includegraphics[scale=0.95]{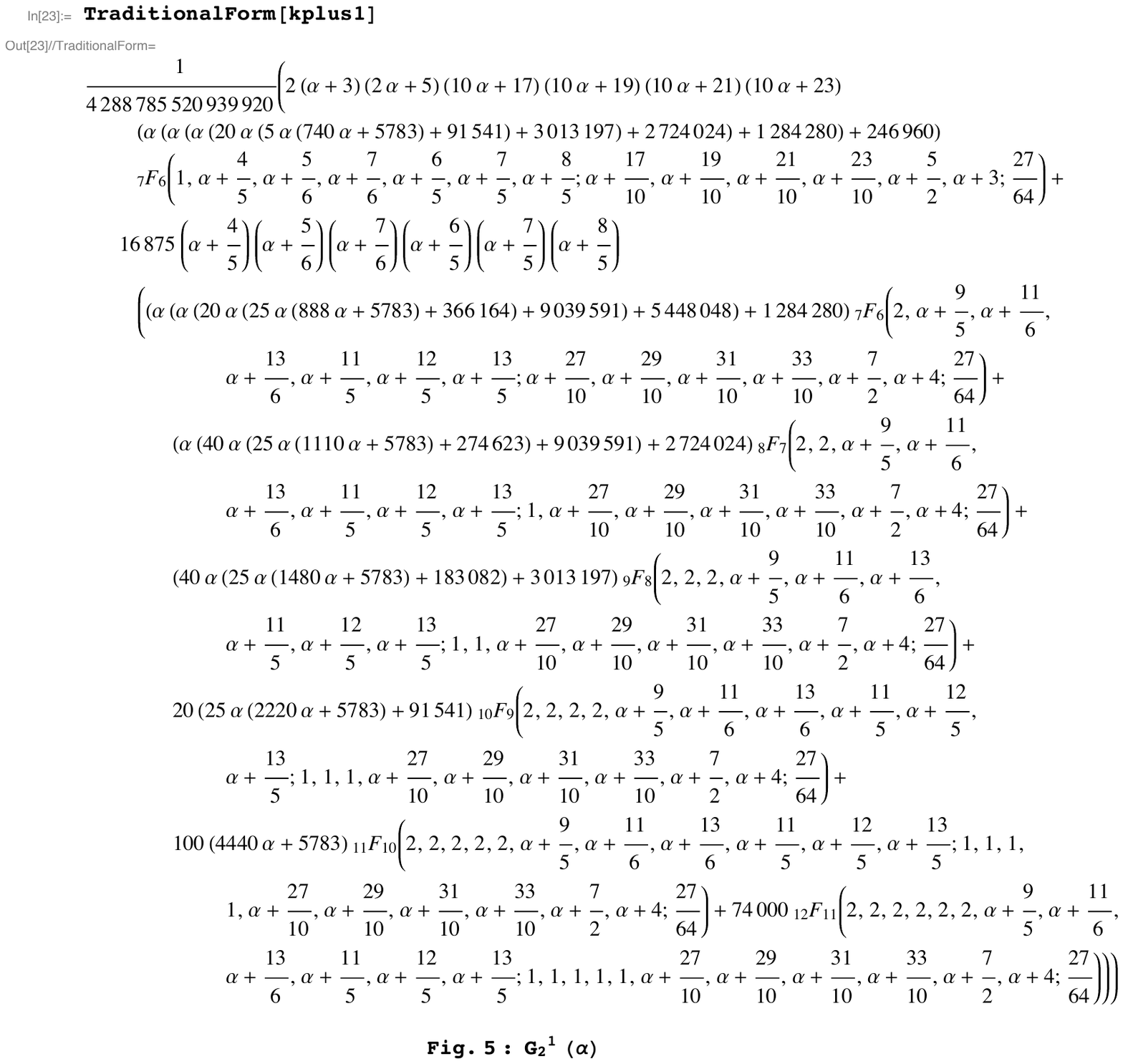}
\caption{\label{fig:kplus1}$G_2^{1}(\alpha)$}
\end{figure}

\begin{figure}
\includegraphics[scale=0.95]{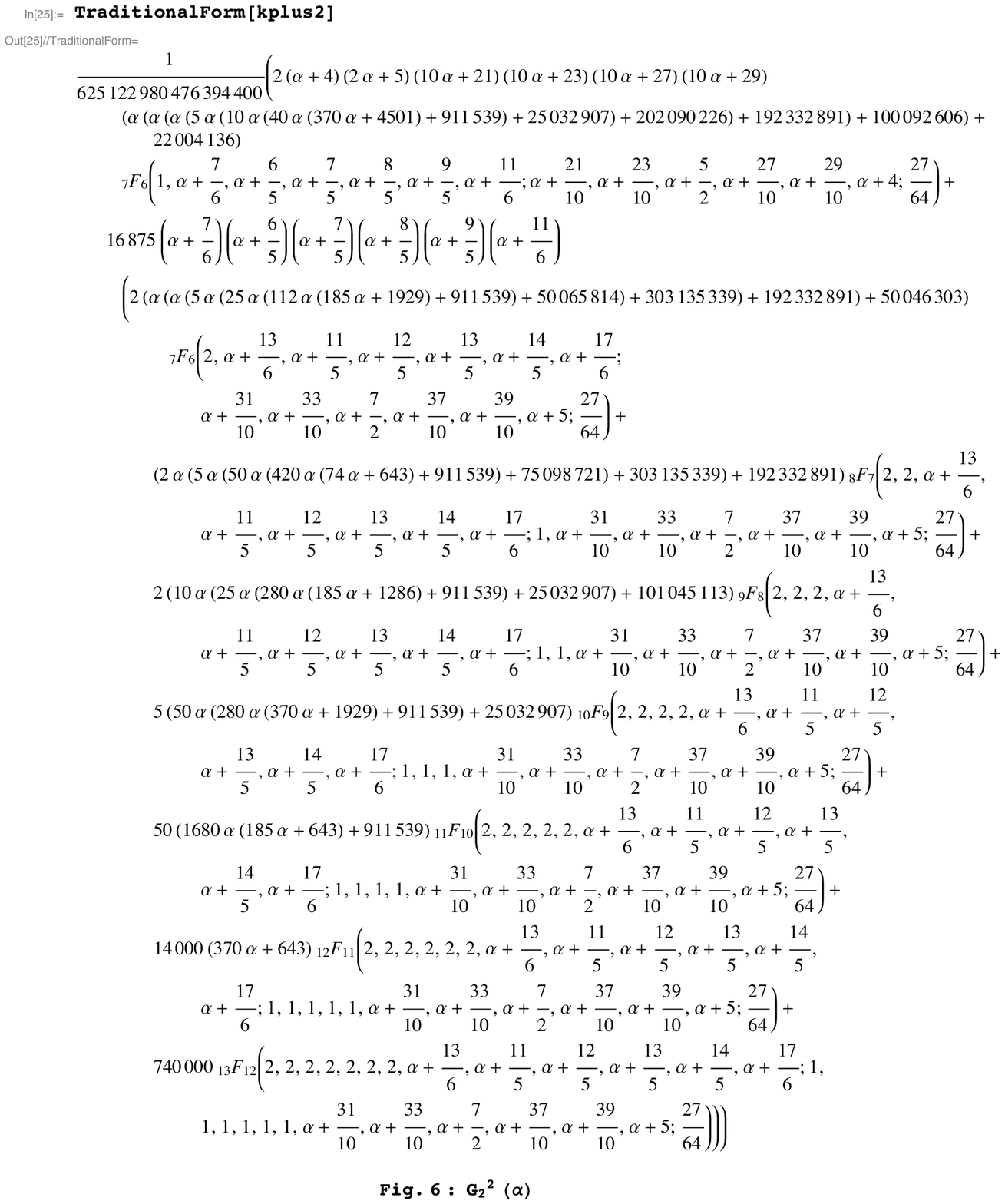}
\caption{\label{fig:kplus2}$G_2^{2}(\alpha)$}
\end{figure}

In Figs.~\ref{fig:kminus1}-\ref{fig:kplus2}, we show a "canonical form" we have 
developed for the factors $G_2^k(\alpha)$ (cf. 
\cite[Fig. 3]{slaterJModPhys}).
\section{Equivalent difference equation forms} \label{Equivalent}
It further appears that all the $G_2^k(\alpha)$ factors ($k = -1,0,1,\ldots,9$)
(Figs. 1-4,...) can be equivalently written as functions that satisfy first-order difference (recurrence) equations of the 
form
\begin{equation}
p_0^k(\alpha) +p_1^k(\alpha) G_2^k(\alpha) +p_2^k(\alpha) G_2^k({1+\alpha}) = 0,
\end{equation}
where the $p$'s are polynomials in $\alpha$ (Figs, 7-12). This was established by yet another application of the Mathematica FindSequenceFunction command. We generated--for each value of $k$ under consideration--a sequence ($\alpha =1,2,\ldots,85$) of the rational values yielded by the hypergeometric-based formulas for $G_2^k(\alpha)$, to which the command was then applied.
\begin{figure}
\includegraphics[scale=0.95]{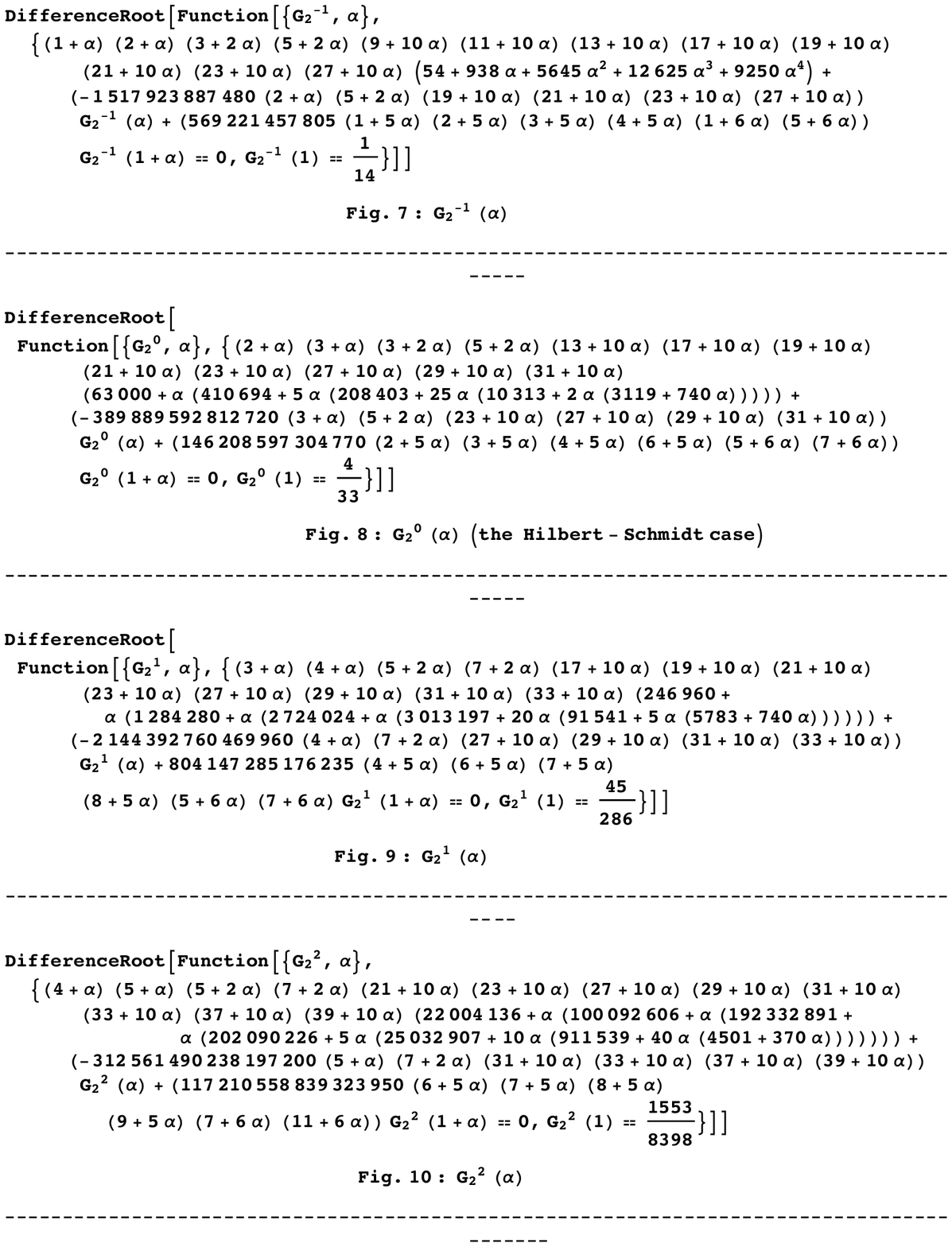}
\caption{\label{fig:DF1}Difference Equation Forms of $G_2^k(\alpha)$ for $k = -1, 0, 1, 2$}
\end{figure}
\begin{figure}
\includegraphics[scale=0.95]{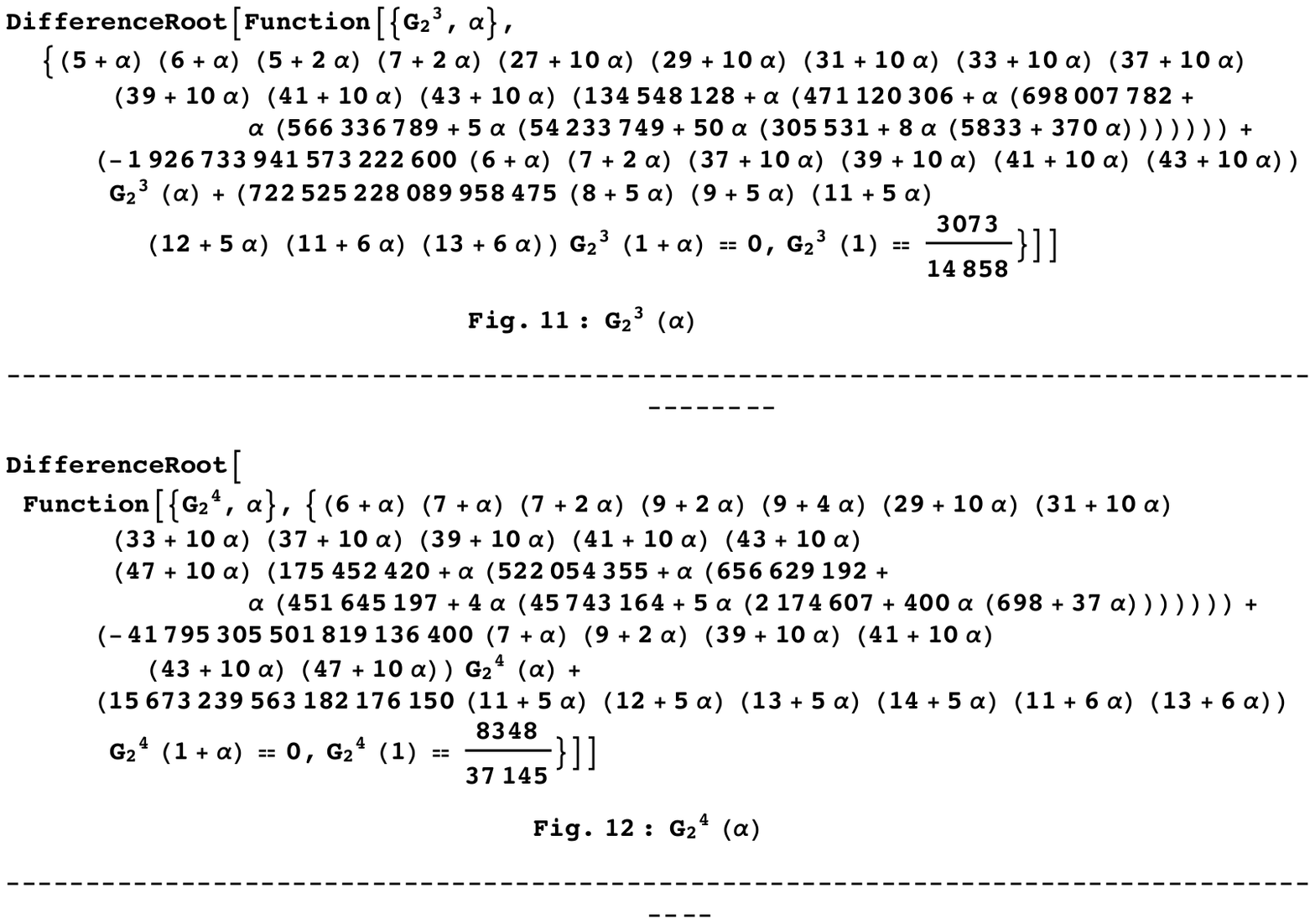}
\caption{\label{fig:DF2}Difference Equation Forms of $G_2^k(\alpha)$ for $k = 3, 4$}
\end{figure}
While we have limited ourselves in the last six figures to displaying our results 
for $k=-1,0,1, 2, 3$ and 4, we do have the analogous set of results in terms of the hypergeometric functions for 
the additional instances, $k=5,6,7,8$ and 9, and presume that an equivalent set of
difference-equation results is constructible (though substantial efforts with $k=5$ have not to this point succeeded).  The initial points $G_2^k(1)$ in the six difference equations shown are--in the indicated order--$\left\{\frac{1}{14},\frac{4}{33},\frac{45}{286},\frac{1553}{8398},\frac{3073}{14858},\frac{8348}{37145}\right\}$. The next five members of the sequence are 
$\left\{\frac{188373}{785726},\frac{1096583}{4342170},\frac{6050627}{22951470},\frac{160298199}{586426690},\frac{13988600951}{49611697974}\right\}$. 
Since $G_1^k(1)=1$, these are the respective separability probabilities $Q(k,1)$ themselves. We would like to extend this sequence sufficiently, so that we might be able to establish an underlying rule for it. (However, since the sequence is increasing in value, the Legendre-polynomial density-approximation procedure converges more slowly as $\alpha$ increases, so our quest seems somewhat problematical, despite the large number of moments incorporated  [cf. \cite[App. II]{LatestCollaboration}].)
If in the difference equation for $k=-1$ (Fig. 7), we replace 
$G_2^{-1}(1)=\frac{1}{14}$ by $G_2^{-1}(1)=0$ , then we can add
\begin{equation}
\frac{\pi  3^{-3 \alpha-5} 4^{3 \alpha+2} 5^{5 \alpha+3} \left(\frac{9}{10}\right)_{\alpha+1}
   \left(\frac{11}{10}\right)_{\alpha+1} \left(\frac{13}{10}\right)_{\alpha+1}
   \left(\frac{3}{2}\right)_{\alpha+1} \left(\frac{17}{10}\right)_{\alpha+1} \Gamma (\alpha) \Gamma
   (\alpha+2)}{52055003 \Gamma (5 \alpha) \Gamma \left(\alpha+\frac{1}{6}\right) \Gamma
   \left(\alpha+\frac{5}{6}\right)},
\end{equation}
to the $\alpha$-specific values obtained from the so-modified equation to recover the values generated by the original $k=-1$ difference equation (Fig. 7).
\subsection{Polynomial coefficients in difference equations}
We have for five  ($k=-1,1,2,3,4$) of the six  cases at hand  (Figs. 7-12) the proportionality
relation
\begin{equation} \label{p2Proportion}
p_2^k(\alpha) \propto \Pi_{i=1}^6 (u_i-1),
\end{equation}
where the $u_i$'s (and $b_i$'s) are themselves functions of both $k$ and $\alpha$.
The (symmetric/Hilbert-Schmidt) case $k=0$ fails to conform to this relationship because a factor of $(1 +5 \alpha)$ is present in the right-hand-side of (\ref{p2Proportion}), rather than $(6 +5 \alpha)$, as in the corresponding difference equation.
Now, for all six displayed cases (including $k=0$),
\begin{equation}
p_1^k(\alpha) \propto \Pi_{i=1}^6 b_i  .
\end{equation}
Further, for all six cases, the polynomial coefficients $p_0^k(\alpha)$ 
are proportional to the product of a factor of the form
\begin{equation}
\Pi_{i=1}^6 b_i (b_i-1),
\end{equation}
and an irreducible polynomial. These polynomials are, in the indicated order,
\begin{equation}
9250 \alpha ^4+12625 \alpha ^3+5645 \alpha ^2+938 \alpha +54,
\end{equation}
\begin{equation}
185000 \alpha ^5+779750 \alpha ^4+1289125 \alpha ^3+1042015 \alpha ^2+410694 \alpha
   +63000,
\end{equation}
\begin{equation}
74000 \alpha ^6+578300 \alpha ^5+1830820 \alpha ^4+3013197 \alpha ^3+2724024 \alpha
   ^2+1284280 \alpha +246960,
\end{equation}
and (for $k=2$)
\begin{equation}
740000 \alpha ^7+9002000 \alpha ^6+45576950 \alpha ^5+125164535 \alpha ^4 +202090226
   \alpha ^3
\end{equation}
\begin{displaymath}
+192332891 \alpha ^2+100092606 \alpha +22004136.
\end{displaymath}
 The irreducible polynomial for $k=3$ is also of degree 7, that is,
\begin{equation}
740000 \alpha ^7+11666000 \alpha ^6+76382750 \alpha ^5+271168745 \alpha ^4+566336789
   \alpha ^3\end{equation}
\begin{displaymath}
+698007782 \alpha ^2+471120306 \alpha +134548128.
\end{displaymath}
For $k=4$, this auxiliary polynomial is now the product of $(9 +4 \alpha)$ times an irreducible  polynomial of degree 7, that is,
\begin{equation}
296000 \alpha ^7+5584000 \alpha ^6+43492140 \alpha ^5+182972656 \alpha ^4+451645197
   \alpha ^3
\end{equation}
\begin{displaymath}
+656629192 \alpha ^2+522054355 \alpha +175452420.
\end{displaymath}
The coefficients of the highest powers of $\alpha$ in all six irreducible polynomials are factorable into the product of 37 and powers of 2 and 5. 
\section{$\mbox{Prob}(|\rho|^{PT}>0)$ Analyses}
Efforts of our to conduct parallel sets of ($k$-specific) analyses to those reported above
for {\it total} separability probabilities ($|\rho^{PT}| > 0$), rather than for that component part
of the probabilities satisfying the determinantal inequality 
$|\rho^{PT}| >
|\rho|$ have so far been unsuccessful, in the following sense. We have computed what appear to be appropriate sequences 
$(\alpha =1, 2,\ldots,74)$ of rational values for $k=1$ 
and $(\alpha =1, 2,\ldots,124)$ for $k=2$, 
but the Mathematica  FindSequenceFunction 
has not produced any underlying governing rules. (This can be contrasted with the results in \cite{LatestCollaboration}, where such  successes were reported in obtaining 
$\alpha$-specific [$|\rho^{PT}| > 0$] formulas [$\alpha = 1,2,\ldots,13$ and $\frac{1}{2}, \frac{3}{2},\frac{5}{2},\frac{7}{2}$].) 

\begin{figure}
\includegraphics{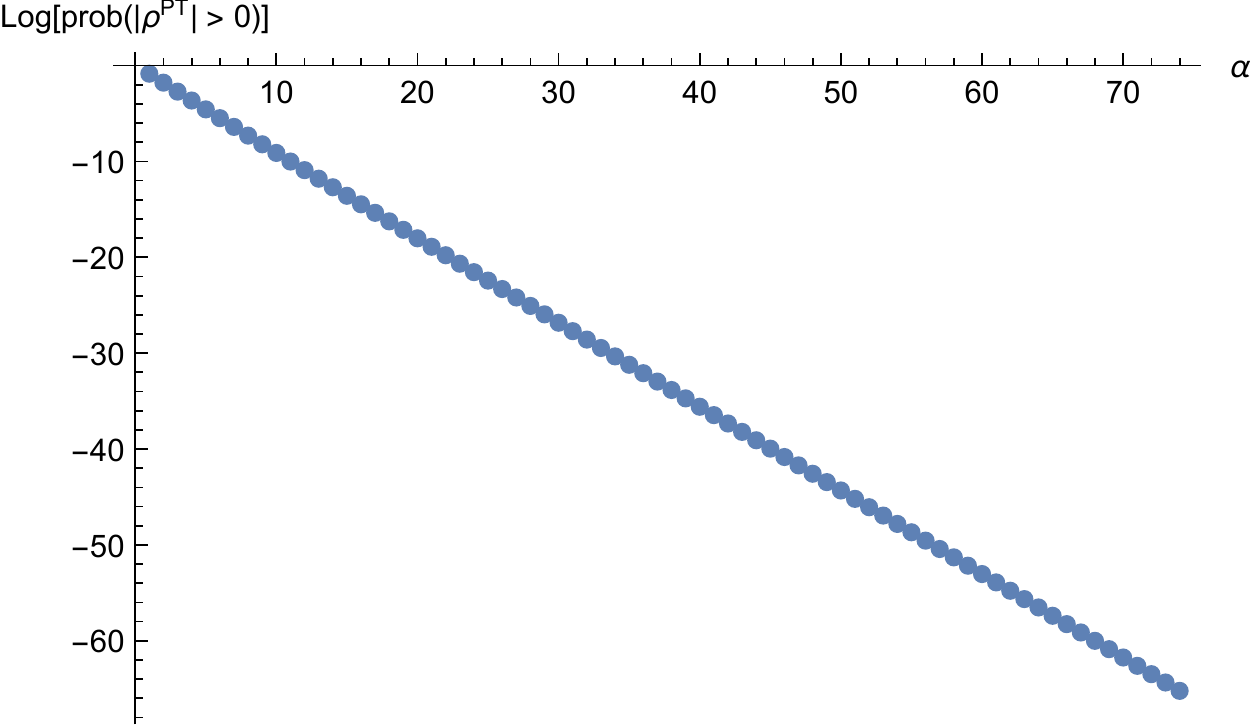}
\caption{\label{fig:Full}Plot of logs of separability probability
($|\rho^{PT}| > 0$) for random induced measure with $k=1$. A least-squares linear fit to these 74 points is $-0.878482 \alpha -0.362781$.}
\end{figure}

\begin{figure}
\includegraphics{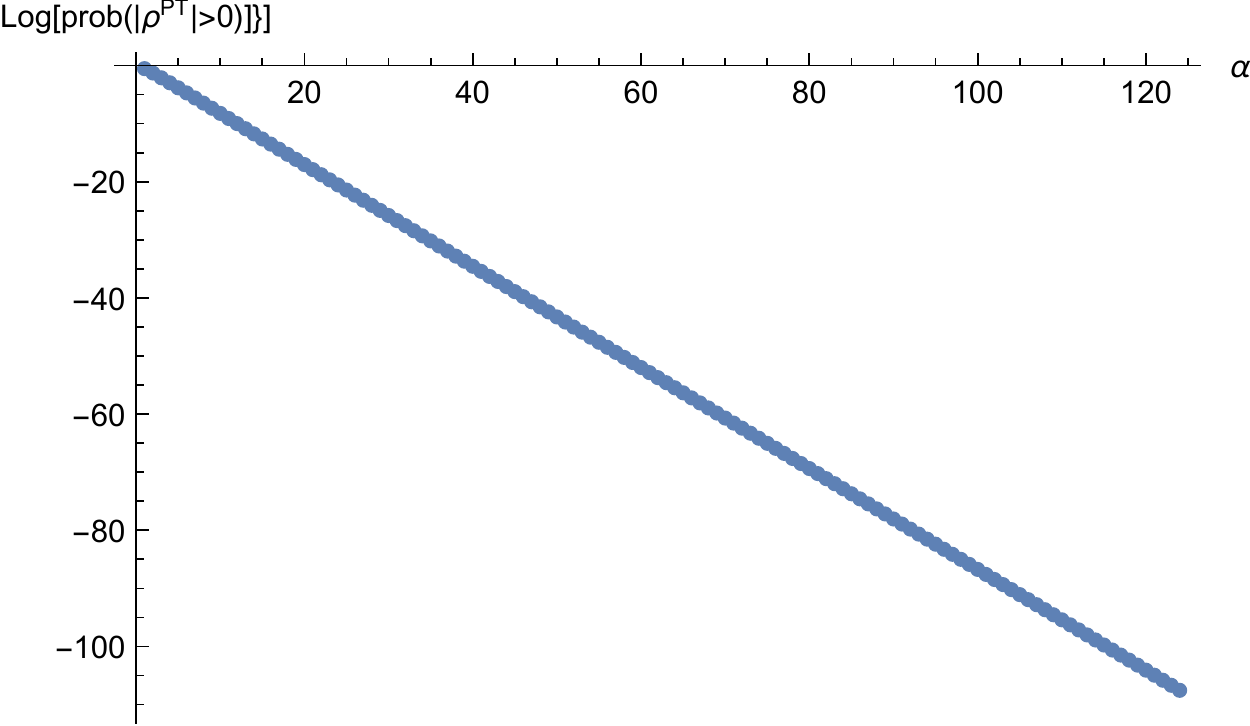}
\caption{\label{fig:Full2}Plot of logs of separability probability
($|\rho^{PT}| > 0$) for random induced measure with $k=2$.  A least-squares linear fit to these 124 points is $-0.871033 \alpha + 0.351201$.}
\end{figure}

In Fig.~\ref{fig:Full}, we plot the logs
of these $k=1$ seventy-four total separability probabilities (based on $\alpha=1,\ldots,74$). A least-squares linear fit to these points is $-0.878482 \alpha -0.362781$, while in Fig.~\ref{fig:Full2}, we show (based on $\alpha=1,\ldots,124$) the $k=2$ counterpart, with an analogous fit of $-0.871033 \alpha + 0.351201$.
Although the slopes of these two linear fits are quite close, the $y$-intercepts themselves are of different sign. The predicted probabilities at $\alpha=1$, the first of the fitted points, are 0.289019 and 0.602955, respectively. In statistical parlance, the "coefficients of determination" or $R^2$ for the two linear fits to the log-plots are both greater than 0.99995. Further, sampling at $\alpha =1, 51, 101,\ldots, 1451$, we obtained an estimated, again, very-well fitting line of $-1.4754 -0.86417 \alpha$.
\subsection{Asymptotic properties}
\subsubsection{$k$-specific $\mbox{prob}(|\rho^{PT}| >0)$ formulas}
Dunkl, on the basis of our $k =1, \alpha =1, 51, 101,\ldots, 1451$ analysis just above (and its companions), did
advance the bold and (certainly, in our overall analytical context) elegant  hypothesis of a $k$-invariant ($\alpha \rightarrow \infty$) slope equal to $\log{\frac{27}{64}} 
\approx -0.8630462173553$, which does seem quite consistent with the numerical properties
we have observed (that is, with the direction in which the estimates of the slope tend as the number of points sampled increase). As further support, we obtained for a  $k =2, \alpha =1, 49, 73 ,\ldots, 1465$ analysis, a slope estimate of -0.864025, again converging in the direction of 
$\log{\frac{27}{64}} $.
(Let us remark,  regarding the generalized two-qubit version of the [simpler, lower-dimensional] X-states model \cite{Xstates2,LatestCollaboration2,Beran}, that C. Dunkl has been able to show that the slope of a [now, log-log] plot 
of $\log({\mbox{prob}(|\rho^{PT}|>0})$ vs. $\log{\alpha}$ tends to $-\frac{1}{2}$, as $\alpha \rightarrow \infty$.)
\subsubsection{$\alpha$-specific $\mbox{prob}(|\rho^{PT}| >0)$ formulas}
\begin{figure}
\includegraphics{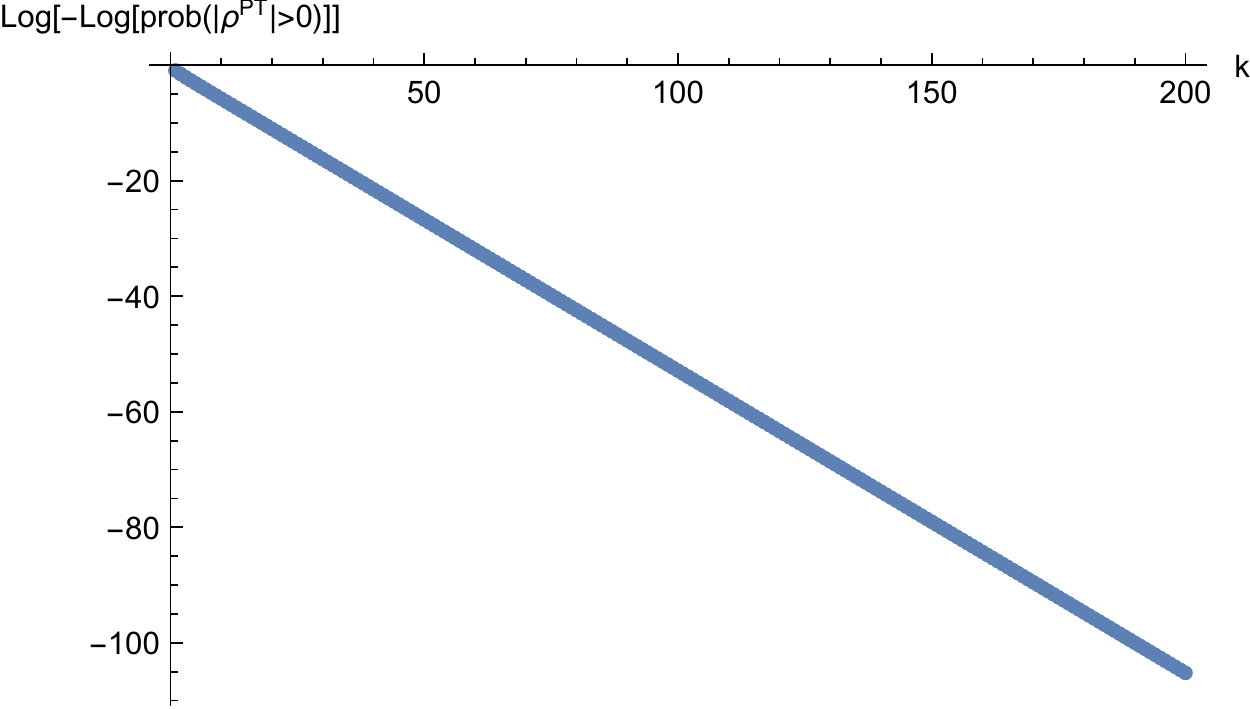}
\caption{\label{fig:Dual}Plot of $\log({-(\log{P^{rebit}_k}}))$  vs. $k$. The slope of a least-squares-fitted line is 
-0.523280, while $\log{\frac{16}{27}} \approx -0.523248$.}
\end{figure}

These interesting observations led us to reexamine, for their asymptotic properties, the "dual" $\alpha$-specific formulas reported in \cite{LatestCollaboration}.
We now find--through analytic means--that for each of $\alpha = 1,2,3,4$ and $\frac{1}{2}, \frac{3}{2},\frac{5}{2},\frac{9}{2}$, that as $k \rightarrow \infty$, the ratio of the logarithm of the 
$(k+1)$-st separability probability to the logarithm of the 
$k$-th separability probability is $\frac{16}{27}$. (Presumably, the pattern continues for larger $\alpha$, but the required computations have, so far, proved too challenging.)
For example, for $\alpha =\frac{1}{2}$, we have for the two-rebit total separability probability, as a function of $k$, the formula
\cite[eq. (4)]{LatestCollaboration}
\begin{equation}
P^{rebit}_k=1-\frac{4^{k+1} (8 k+15) \Gamma (k+2) \Gamma \left(2
   k+\frac{9}{2}\right)}{\sqrt{\pi } \Gamma (3 k+7)}.
\end{equation}
In Fig.~\ref{fig:Dual}, we show a plot
of $\log({-(\log{P^{rebit}_k))}}$  vs. $k$. The slope of a 
least-squares-fitted line based on the 200 points is 
-0.523280, while $\log{\frac{16}{27}} \approx -0.523248$.
(As we increase $\alpha$ from $\frac{1}{2}$, but hold the number of points constant at 200, the approximation of the slope to this value slowly weakens.)
\subsubsection{$k$-specific $\mbox{prob}(|\rho^{PT}| >|\rho|)$ formulas}
Now, as concerns the eleven ($k =-1,0,1,\ldots,9$) formulas for $\mbox{prob}(|\rho^{PT}| >|\rho|)$, which have been the principal focus of the paper, we have computed the ratios of the probability for  $\alpha=101$ to the 
probability for $\alpha =100$. These ranged from 0.419810 ($k=-1$) to 0.4204296 ($k=9$).
Let us note here that $\frac{27}{64} \approx 0.421875$.
\subsubsection{$\alpha$-specific $\mbox{prob}(|\rho^{PT}| >|\rho|)$ formulas}
We had available $\alpha =\frac{1}{2}, 1$ and 2 computations for $k=1,\ldots,40$ for this scenario. 
We found that, for each of the three values of $\alpha$, we could construct strongly linear plots--with unit-like slopes between 1.00177 and 1.00297--by
taking $k$ times the ratio ($R$) of the $(k+1)$ separability probability to the $k$-th separability probability. (From this, it appears, simply, that $R \rightarrow 1$, as 
$k \rightarrow \infty$.)
\subsubsection{"Diagonal" $\alpha=k$ $\mbox{prob}(|\rho^{PT}| >|\rho|)$ formulas}
For values $\alpha =k =1,\ldots,50$, we were able to construct
a strongly linear plot by--similarly to the immediate last analysis--taking $k=\alpha$ times the ratio of the $(k+1)=(\alpha+1)$ separability probability to the $k=\alpha$-th separability probability. Now, however, rather than a slope very close to 1, we found a slope near to one-half, that is 0.486882. The ($k=\alpha =0$)-intercept of the estimated line was 0.894491.
\section{"Concise formulas"}
Let us also remind the reader of the interesting "concise" (Hilbert-Schmidt [$k=0$]) generalized two-qubit result--applying Zeilberger's ("telescoping") algorithm \cite{doron}--of Qing-Hu Hou, reported in \cite[eqs. (1)-(3)]{slaterJModPhys}.
This--in our present notation--takes the form (cf. Figs. 5, 9)
\begin{equation} \label{Hou1}
Q(0,\alpha) =\Sigma_{i=0}^\infty f(\alpha+i),
\end{equation}
where
\begin{equation} \label{Hou2}
f(\alpha) = Q(0,\alpha)-Q(0,\alpha +1) = \frac{ q(\alpha) 2^{-4 \alpha -6} \Gamma{(3 \alpha +\frac{5}{2})} \Gamma{(5 \alpha +2})}{6 \Gamma{(\alpha +1)} \Gamma{(2 \alpha +3)} 
\Gamma{(5 \alpha +\frac{13}{2})}},
\end{equation}
and
\begin{equation} \label{Hou3}
q(\alpha) = 185000 \alpha ^5+779750 \alpha ^4+1289125 \alpha ^3+1042015 \alpha ^2+410694 \alpha +63000 = 
\end{equation}
\begin{displaymath}
\alpha  \bigg(5 \alpha  \Big(25 \alpha  \big(2 \alpha  (740 \alpha
   +3119)+10313\big)+208403\Big)+410694\bigg)+63000.
\end{displaymath}
We divide the originally reported formula by one-half, since we have moved here from the ($k=0$) Hilbert-Schmidt $|\rho^{PT}| >0$ original 
scenario to its  $|\rho^{PT}| > |\rho|$ counterpart.
Using our earlier results above, Hou has been able to construct the $k=1$ analogue of the
"concise formula",
\begin{equation} \label{k1Hou1}
Q(1,\alpha) =\Sigma_{i=0}^\infty f(\alpha+i),
\end{equation}
where
\begin{equation}
f(\alpha)=  \frac{q(\alpha) \left(27 \right)^{\alpha } \Gamma (5 \alpha ) \Gamma \left(\alpha
   +\frac{5}{6}\right) \Gamma \left(\alpha +\frac{7}{6}\right)}{\left(50000 \right)^{\alpha }\Gamma (\alpha ) \Gamma
   \left(\alpha +\frac{17}{10}\right) \Gamma \left(\alpha +\frac{19}{10}\right) \Gamma
   \left(\alpha +\frac{21}{10}\right) \Gamma \left(\alpha +\frac{23}{10}\right) \Gamma
   (2 \alpha +5)}
\end{equation}
and
\begin{equation}
q(\alpha)=\frac{9 \pi}{1000000} (5 \alpha +1) (5 \alpha +2) (5 \alpha +3) \times
\end{equation}
\begin{displaymath}
 \left(74000 \alpha ^6+578300 \alpha ^5+1830820
   \alpha ^4+3013197 \alpha ^3+2724024 \alpha ^2+1284280 \alpha +246960\right).
\end{displaymath}

\begin{acknowledgments}
PBS expresses appreciation to the Kavli Institute for Theoretical
Physics (KITP) for computational support in this research and, of course, to 
Charles Dunkl for his many, many expert contributions and interactions in this research program in the past few years. Qing-Hu Hou has, as indicated in the final section of the paper, been very helpful also.
\end{acknowledgments}

\bibliography{DifferenceHyper2}

\end{document}